\documentclass[12pt]{spieman}  
\usepackage{amsmath,amsfonts,amssymb}
\usepackage{graphicx}
\usepackage{setspace}
\usepackage{tocloft}
\usepackage{rotating}
\usepackage{wrapfig}

\title{The X-ray Polarization Probe mission concept}

\author[a]{Keith Jahoda}
\author[b]{Henric Krawczynski}
\author[c]{Fabian Kislat}
\author[d]{Herman Marshall}
\author[a]{Takashi Okajima}
\affil[a]{NASA/Goddard Space Flight Center, Greenbelt MD, 20771;
$^b$Washington University, St.Louis, MO, 63130;
$^c$University of New Hampshire, Durham, NH, 03824;
$^d$MIT Kavli Institute, Cambridge, MA, 02139}

\cftpagenumbersoff{figure}
\cftpagenumbersoff{table} 
\begin{document} 
\maketitle
\pagenumbering{goggle}

\vspace*{-0.5cm}
Co-authors: Ivan Agudo (CSIC),
Lorella Angelini (GSFC),
Matteo Bachetti (INAF),
Luca Baldini (U. Pisa),
Matthew G. Baring (Rice U.),
Wayne Baumgartner (MSFC),
Ronaldo Bellazzini (INFN),
Stefano Bianchi (U. Roma Tre),
Niccol\`o Bucciantini (INAF/OAC),
Ilaria Caiazzo (UBC),
Fiamma Capitanio (INAF/IAPS),
Paolo Coppi (Yale U.),
Enrico Costa (INAF/IAPS),
Ettore Del Monte (INAF),
Alessandra De Rosa (IAPS/INAF),
Jason Dexter (MPE),
Laura Di Gesu (ASI),
Niccolo' Di Lalla (INFN),
Victor Doroshenko (U. T\"ubingen),
Michal Dovciak (Cz. Acad. Sci.),
Riccardo Ferrazzoli (INAF/IAPS),
Felix F\"urst (ESAC),
Alan Garner (MIT),
Pranab Ghosh (Tata Inst.),
Denis Gonz\'alez-Caniulef (UBC),
Victoria Grinberg (U. T\"ubingen),
Shuichi Gunji (Yamagata U.),
Dieter Hartmann (Clemson U.),
Kiyoshi Hayashida (Osaka),
Jeremy Heyl (UBC),
Joanne Hill (GSFC),
Adam Ingram (Oxford U.),
Wataru Buz Iwakiri (Chuo U.),
Svetlana Jorstad (BU),
Phil Kaaret (U. Iowa),
Timothy Kallman (GSFC),
Vladimir Karas (Cz. Acad. Sci.),
Ildar Khabibullin (MPA),
Takao Kitaguchi (RIKEN),
Jeff Kolodziejczak (MSFC),
Chryssa Kouveliotou (GWU),
Ioannis Liodakis (Stanford U.),
Thomas Maccarone (Texas Tech U.),
Alberto Manfreda (INFN),
Frederic Marin (U. Strasbourg),
Andrea Marinucci (ASI),
Craig Markwardt (GSFC),
Alan Marscher (BU),
Giorgio Matt (U. Roma Tre),
Mark McConnell (UNH),
Jon Miller (U. Michigan),
Ikuyuki Mitsubishi (Nagoya U.),
Tsunefumi Mizuno (U. Hiroshima),
Alexander Mushtukov (Leiden U.),
C.-Y. Ng (Hong Kong U.),
Michael Nowak (Washington U.),
Steve O'Dell (MSFC),
Alessandro Papitto (INAF/OAR),
Dheeraj Pasham (MIT),
Mark Pearce (KTH),
Lawrence Peirson (Stanford U.),
Matteo Perri (SSDC/ASI),
Melissa Pesce-Rollins (INFN),
Vahe Petrosian (Stanford U.),
Pierre-Olivier Petrucci (U. Grenoble),
Maura Pilia (INAF/OAC),
Andrea Possenti (INAF/OAC, U. Cagliari),
Juri Poutanen (U. Turku),
Chanda Prescod-Weinstein (UNH),
Simonetta Puccetti (ASI),
Tuomo Salmi (U. Turku),
Kevin Shi (MIT),
Paolo Soffitta (IAPS/INAF),
Gloria Spandre (INFN),
James F. Steiner (SAO),
Tod Strohmayer (GSFC),
Valery Suleimanov (U. T\"ubingen),
Jiri Svoboda (Cz. Acad. Sci.),
Jean Swank (GSFC),
Toru Tamagawa (RIKEN),
Hiromitsu Takahashi (Hiroshima U.),
Roberto Taverna (U. Roma Tre),
John Tomsick (UCB),
Alessio Trois (INAF/OAC),
Sergey Tsygankov (U. Turku),
Roberto Turolla (U. Padova),
Jacco Vink (U. Amsterdam),
J\"orn Wilms (U. Erlangen-Nuremberg),
Kinwah Wu (MSSL, UCL),
Fei Xie (INAF),
George Younes (GWU),
Alessandra Zaino (U. Roma Tre),
Anna Zajczyk (GSFC, UMBC),
Silvia Zane (MSSL, UCL),
Andrzej Zdziarski (NCAC),
Haocheng Zhang (Purdue U.),
Wenda Zhang (Cz. Acad. Sci.),
Ping Zhou (U. Amsterdam)\\[1ex]

{\bf Abstract:} The {\it X-ray Polarization Probe (XPP)} is a second generation X-ray polarimeter following up on the {\it Imaging X-ray Polarimetry Explorer (IXPE).}
The {\it XPP} will offer true broadband polarimetery over the wide 0.2-60 keV bandpass in addition to imaging polarimetry from 2-8 keV.
The  extended energy bandpass and improvements in sensitivity will enable the simultaneous measurement of the polarization of several emission components. These measurements will give qualitatively new information about how compact objects work, and will probe fundamental physics, i.e. strong-field quantum electrodynamics and strong gravity.


\pagenumbering{arabic}
\setcounter{page}{1}


\begin{spacing}{1}

\section{Key Science Goals and Objectives}
\label{sect:intro}  
The {\it X-ray Polarization Probe (XPP)} is a mission concept for a second-generation polarimetry mission following up on the {\it Imaging X-ray Polarimetry Explorer (IXPE)} mission to be launched in 2021 \cite{Weisskopf16,Odell18}. 
{\it IXPE} will open the field of observational X-ray polarimetry with a modest instrument, compatible with the constraints of NASA's Small Explorer program, and consequently has a modest bandpass (a factor of ~4 between the limiting energies at the upper and lower ends of the band) and effective area (peaking near 130 cm$^2$ at 2.2 keV). {\it XPP}, as a second generation instrument, aspires to at least an order of magnitude increase in both effective area and band width and to substantially improved imaging capabilities. {\it XPP} will extend the 2-8 keV bandpass 
of {\it IXPE} to 0.2 keV - 60 keV and  improve on {\it IXPE's} sensitivity in {\it IXPE's} core energy range from 2-8 keV by a factor between 3 and 10. The {\it XPP} includes two telescopes with {\it Hitomi} style mirrors simultaneously illuminating three instruments which span the 0.2 - 60 keV band and one telescope with {\it IXPE}-type imaging polarimetric capabilities (15'' Half Power Diameter, HPD) 
in the 2-8 keV energy range. 
The use of novel Si mirrors may make it possible to improve the angular resolution of all three telescopes to a few arc sec.

These capabilities will make it possible to obtain quantitatively new information about the most extreme objects in the Universe: black holes, neutron stars, magnetars and Active Galactic Nuclei. Although the number of objects accessible to {\it XPP} is still modest (a few hundred sources with high signal to noise ratio polarization measurements), the observations will enable physics-type experiments probing the inner workings of these sources of high-energy X-rays, and probing the underlying physical laws in truly extreme conditions. Our {\it XPP} science white
paper\cite{Krawczynski19} identified three  high-profile science investigations: (1) Dissect the structure of inner accretion flow onto black holes and observe strong gravity effects; (2) Use neutron stars as fundamental physics laboratories; (3) Probe how cosmic particle accelerators work and what role  magnetic fields play. 

\begin{figure}[t]
\begin{center}
\begin{tabular}{c}
\includegraphics[height=5.5cm]{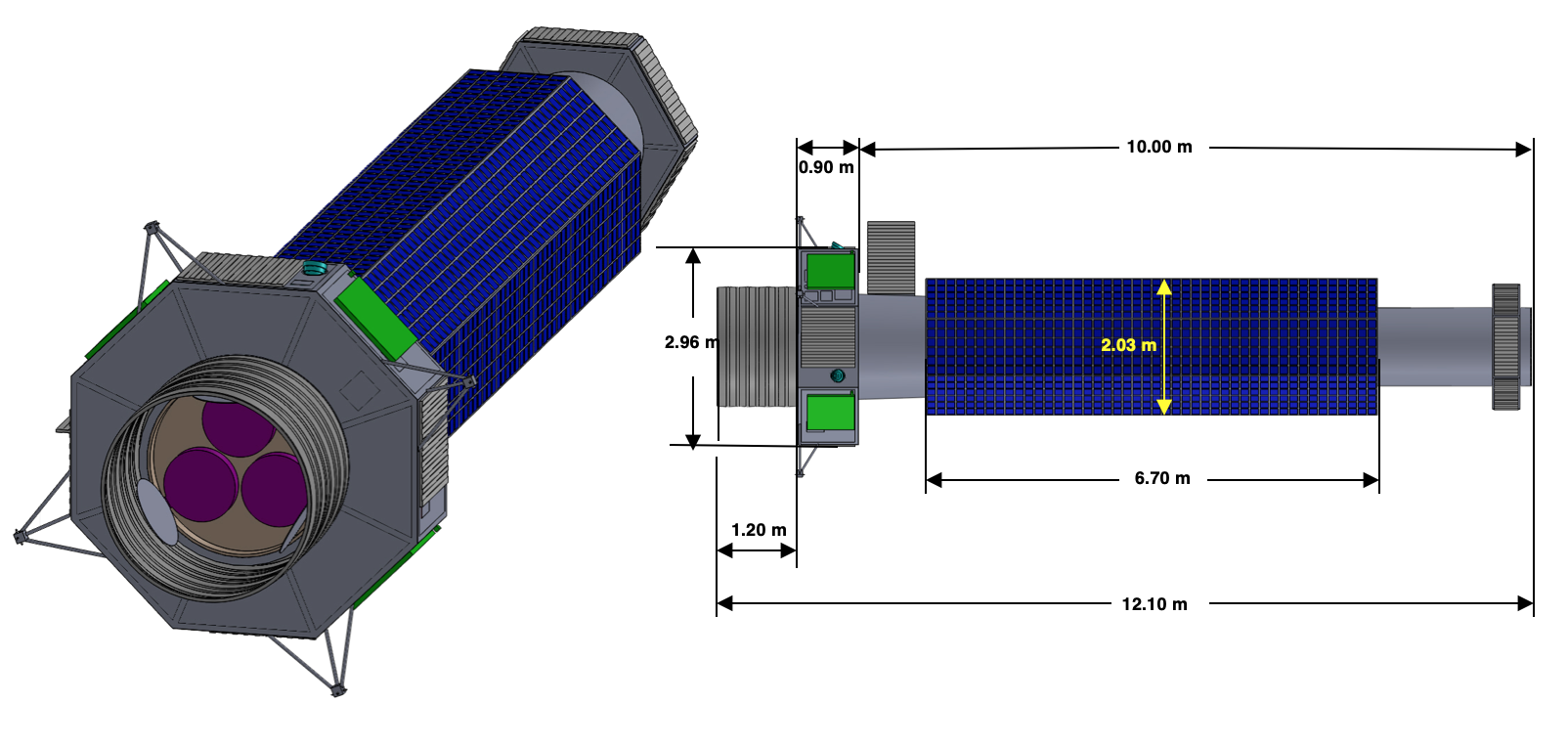}
\end{tabular}
\end{center}
\caption 
{ \label{fig:mission}
{\it XPP} employs fixed solar panels surrounding a 10 m optical bench and fixed phased-array antennas for communication.  The only (one-time) deployables are a sun shade and the three telescope doors.} 
\end{figure}

Here we report on a possible implementation of the {\it XPP} based on a detailed mission implementation study performed in cooperation with 
the COMPASS Team at the NASA Glenn Research Center. The {\it XPP} science objectives require sensitivity to faint signals, which drives requirements for large collecting area over a wide band and precise control of systematic effects. The large area requires grazing incidence mirrors.  The need to limit  systematics leads to tight requirements on alignment and pointing, which {\it XPP} addresses with a stiff and thermally stable optical bench.   The optical bench is designed
without the need for in-orbit extension as shown in Figure \ref{fig:mission}.  To reduce residual and uncalibrated asymmetries in the payload, {\it XPP} will rotate about the line of sight.

\section{Technical Overview}

\subsection{Key Observables}
The {\it XPP} measures the arrival times, energies, and linear polarization of 0.2-60 keV X-rays making use of three complimentary techniques. The instruments exploit polarization dependent scattering cross sections at low energies (below 1 keV), measure the polarization dependent initial direction of the photoelectron that is ejected after a photoelectric absorption at intermediate 2-10 keV photon energies, and measure the polarization dependent direction of a Compton scattered photon at 10-60 keV photon energies.  The cross over in energy between photoelectric dominated interactions and Compton dominated interactions depends on the atomic number of the absorbing medium;  for the {\it XPP} instrument complement, the  photoelectric process dominates the sensitivity below $\sim$10 keV and the Compton process dominates above.    Separate photon counting instruments observe simultaneously below 1~keV (scattering), below 10~keV (photoelectric), and above 10~keV (Compton).

\subsection{Instrument and Spacecraft Performance Requirements}
The {\it XPP} measures the arrival times and energies of 0.2-60 keV X-rays with $<1{\mu}$s and $<20$\% energy resolution at all wavelengths, respectively. Most importantly, 
it measures the linear polarization fraction and 
polarization angle. The design aims at a sensitivity of $<$1\% Minimum Detectable Polarization (MDP) at 99\%
confidence level for a 10$^5$ sec observation
of a 1 mCrab source. This sensitivity will enable the 
study of statistical samples (each: $\sim$10-50 sources) of the key sources: stellar mass black holes, active galactic nuclei, pulsars, magnetars, X-ray binaries, super nova remnants, and blazars. Systematic errors below 0.2\% in the 
polarization fraction are required to measure the typical polarization fractions of $\sim$5\% with sufficient accuracy to distinguish between competing models. Imaging polarimetry with angular resolutions between 1'' (goal) and 15'' (requirement) enable the studies of extended sources such as pulsar wind nebulae, supernova remnants, and the jets from active galactic nuclei.

\begin{figure}
\begin{center}
\begin{tabular}{c}
\includegraphics[height=6cm]{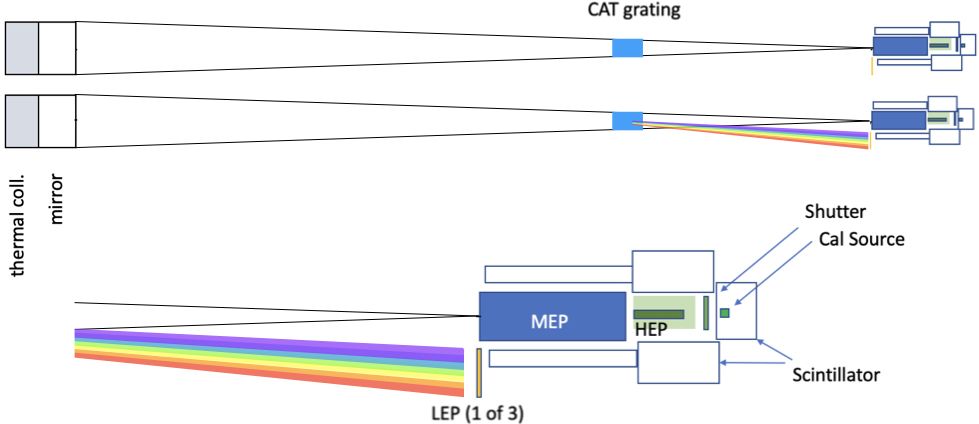}
\end{tabular}
\end{center}
\caption 
{ \label{fig:layout}
The top sketch shows the instrument layout in cross section for the spectro-polarimeter telescopes.   Gratings which disperse low energy X-rays to the Low Energy Polarimeter (LEP) are located 3 m in front of the focal planes.  The bottom sketch shows an expanded view of the photo-electric Medium Energy Polarimeter (MEP)  which sits just in front of the Compton scattering High Energy Polarimeter (HEP).  Each instrument is largely transparent to the energies to which the succeeding instruments are sensitive.} 
\end{figure}

Three of the instruments on {\it XPP} do not have an imaging capability, which creates two high level Attitude Control requirements.   First, pointing must be controlled to a small fraction of the telescope point spread function and second, the Observatory must rotate about the line of sight to both measure and eliminate residual asymmetries in the angular response.

\subsection{Mission Architecture}
The {\it XPP} payload consists of three grazing incidence X-ray mirrors and four focal instruments.   Each mirror has a diameter of 60 cm and a focal length of 10 m.

Two of the telescopes are designed for maximum throughput, and simultaneously illuminate a Low Energy Polarimeter (LEP, sensitive from $\sim$0.2-0.8 keV),  a Medium Energy Polarimeter (MEP, sensitive from $\sim$2 - 10 keV), and a High Energy Polarimeter (HEP, sensitive from $\sim$6 - 60 keV).   These instruments are not imaging, but provide broad band spectral coverage of unconfused point sources.   Most compact object X-ray sources are unconfused and provide excellent opportunities to study matter under extreme conditions.

The third telescope is optimized for imaging, and has a replicated nickel mirror illuminating an imaging polarimeter.  Both the mirror and detector are larger versions of the instruments being developed for {\it IXPE}, and provide imaging polarimetry in the 2-8 keV band.   Imaging confirms that point sources are unconfused, and opens the possibility of studying extended sources such as supernova remnants and jets.

For a launch in the late 2020s, it is likely possible to replace all three mirrors with the multi-layer coated crystalline Silicon optics\cite{Zhang18} being developed for {\it Lynx} and other potential X-ray observatories.  The Si optics are expected to have similar mass and effective area as the {\it Hitomi}-style mirrors and much better angular resolution (near 1'') than the {\it IXPE} mirrors.

The Observatory is put into a deep (7000 x 114000 km), 48 hour orbit which provides over 40 hours of uninterrupted observations of most of the sky (targets  within $30^\circ$ of the sun and anti-sun are unavailable). Data downlink and most other spacecraft functions are performed near perigee.

\begin{figure}
\begin{center}
\begin{tabular}{c}
\includegraphics[height=8.0cm]{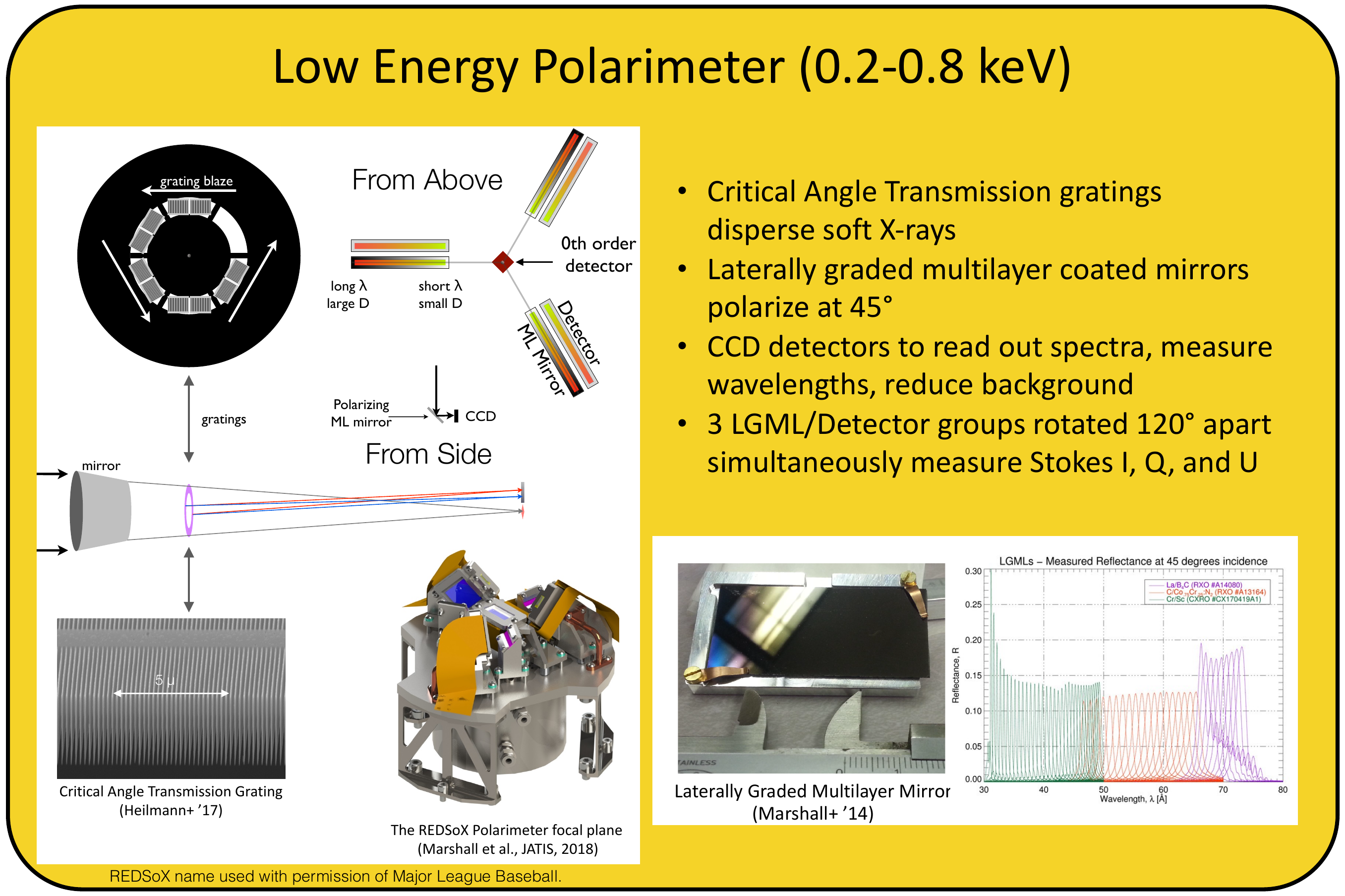}
\end{tabular}
\end{center}
\caption 
{ \label{fig:LEP}
The Low Energy Polarimeter measures the linear polarization at three different angles, enabling a complete determination of the Stokes I, Q, and U parameters.   Observatory rotation will remove any residual uncalibrated asymmetries between the three measurements.} 
\end{figure}

\subsection{Payload Components}

\subsubsection{Spectro-polarimetry telescopes}

The COMPASS design study assumed that two of the mirrors  are optimized for throughput and employ the design recently employed for {\it Hitomi}\cite{Okajima16} but with multi-layer coatings to increase effective area above 10 keV.     Each mirror is 60 cm in diameter and has a 10 m focal length. 

The foil mirrors are expected to have a Half Power Diameter of $\approx 1$ arcmin and the multi-layer reflectors provide about 1700 cm{$^2$} below 10 keV and over 100 cm{$^2$} at all energies above 50 keV.   

Each of these mirrors simultaneously illuminates a Low Energy Polarimeter (LEP), Medium Energy Polarimeter (MEP), and High Energy Polarimeter (HEP) as shown schematically in Figure~\ref{fig:layout}.  The LEP consists of gratings which select energy, variable spacing multilayer mirrors which are tuned to the Bragg angle where they are efficient polarizers and imaging detectors which map detector position to photon energy;  the concept is described by Marshall et al.\cite{Marshall18}.   The gratings are transparent above 2 keV, and their support structure obscures $<$ 10\% of the beam, while scattering low energy photons to gratings/detectors outside the focal plane.  The MEP is a photoelectric polarimeter with a Time Projection Chamber readout\cite{Hill16} based on the detector developed for PRAXyS\cite{Jahoda16}.   The MEP is designed to have both an entrance and an exit window, so that higher energy photons can pass through the MEP and illuminate the HEP.   The HEP is a scattering polarimeter based on the PolSTAR instrument\cite{Krawczynski16} (designed for space operation) and the X-Calibur\cite{Kislat18} balloon payload instrument.    The HEP is surrounded with scintillation detectors to provide an anti-coincidence based background reduction.   A detailed instrument design will be necessary in order to package all three instruments and the associated electronics.   A novel feature of these telescopes is that all three instruments operate simultaneously;  there is no need for a mechanism to move each instrument separately and sequentially into the focal plane.

\begin{figure}
\begin{center}
\begin{tabular}{c}
\includegraphics[height=8.0cm]{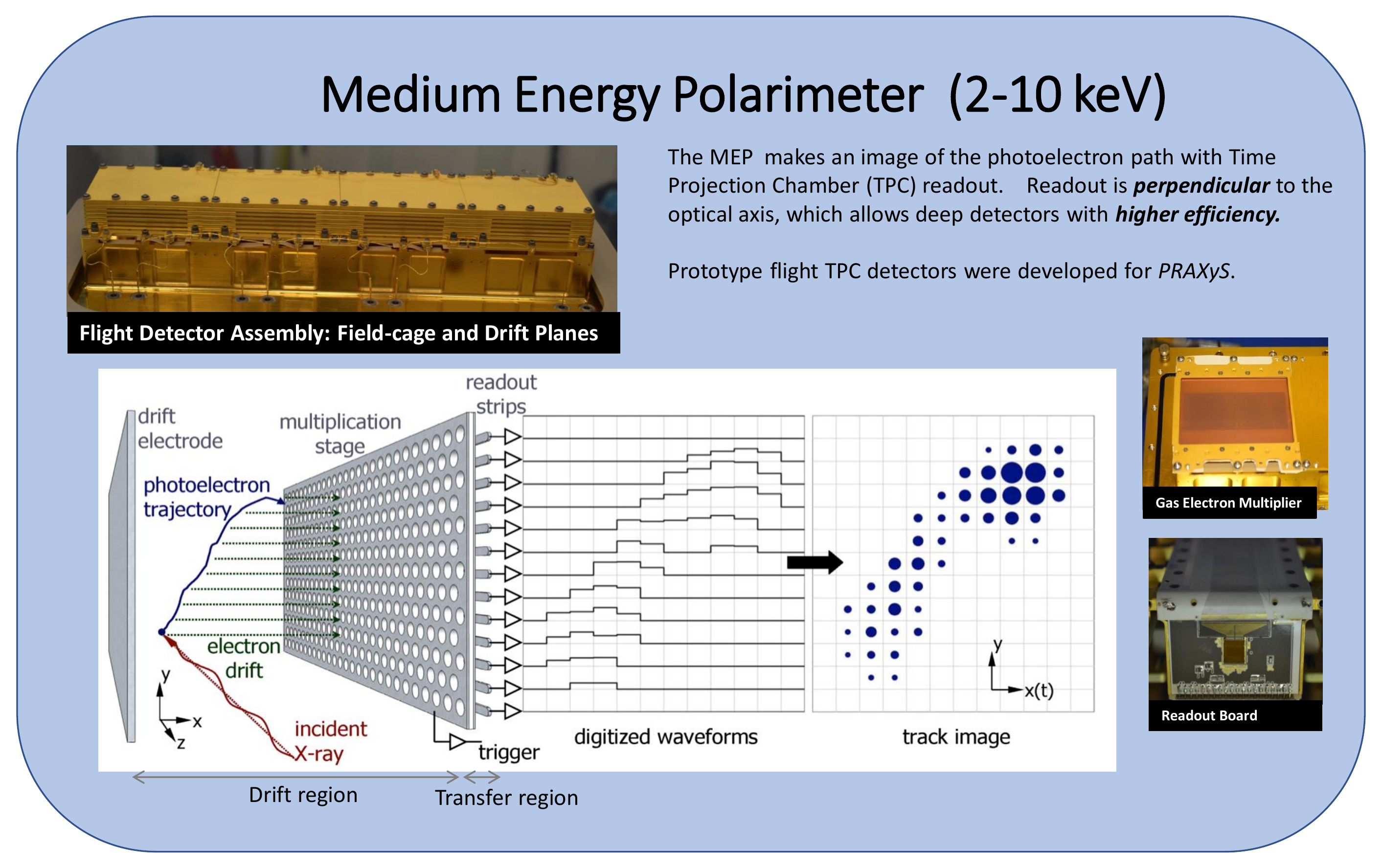}
\end{tabular}
\end{center}
\caption 
{ \label{fig:MEP}
The Medium Energy Polarimeter is a Time Projection Chamber which records an image of the photoelectron track associated with the absorption of a photon in the active volume.} 
\end{figure}

\subsubsection{Imaging polarimetry telescope}
The third mirror has similar diameter and focal length and provides higher angular resolution;  the replicated nickel process  for the IXPE mirrors is described by Ramsey\cite{Ramsey17}.  This mirror illuminates an imaging photoelectric pixel polarimeter\cite{Muleri16} similar to the detector planned for the {\it IXPE}\cite{Weisskopf16}.
This mirror achieves an HPD of 15", and enables polarimetric imaging at sub arc minute scales.

\subsection{Focal Plane Instruments}
Unlike the instruments in the pioneering sounding rocket \cite{Novick72} and satellite \cite{Weisskopf78} experiments, which {\it required} rotation about the line of sight to develop the polarization signature,
the {\it XPP}  polarimeters are each capable of measuring polarization without rotation.   This allows Observatory rotation to be employed to reduce and remove any systematic effects which arise from azimuthal asymmetries in the mirrors or instruments.

\subsubsection{Low Energy Polarimeter}

The Low Energy Polarimeter (LEP) exploits the polarization dependence of Bragg scattering. Simultaneous measurements are performed at three angles. The LEP is made of three well-studied components.   Transmission gratings separate the incident radiation by wavelength dispersing onto custom laterally graded multilayer coated mirrors
(LGMLs) that serve as polarization sensitive analyzers, reflecting only one linear polarization onto imaging detectors.  The LGMLs are  aligned  and mounted such that each dispersed photon strikes the mirrors at an angle that satisfies the Bragg condition for reflection at about 45$^\circ$.
At this angle, the polarization in the plane of reflection is highly suppressed, yielding modulation factors over 90\%.   The detector coordinates give the photon wavelength, as in other dispersive
spectrometers.  With CCDs as readout detectors, the spectral resolution assists in background reduction as the dispersion determines the energy of interest.   More conceptual detail is provided in Marshall et al.\ (2018)\cite{Marshall18}.  Using LGMLs at three distinct orientations (e.g., 120$^\circ$ to each other) provides enough information to measure the I, Q, and U Stokes parameters from which the linear polarization angle and magnitude can be determined.  

Gratings closer to the mirror provide greater dispersion (and thus potentially more precision in the energy determination), at the cost of requiring larger grating area and larger detectors.
Critical angle transmission gratings designed for {\it ARCUS}\cite{2015SPIE.9603E..14H}
with 200 nm periods and LGMLs with period variations
of 0.88 \AA/mm\cite{2015SPIE.9603E..19M} have been tested in the lab and are sufficient for this project.
For the current design, we assume that the gratings are 3 m above the focal plane;  a 0.2 keV photon is dispersed 93 mm while a 0.7 keV photon is dispersed 26.5 mm.    As the optical bench is fixed rather than extensible, this distance could be modified without affecting the conceptual design.

\begin{figure}
\begin{center}
\begin{tabular}{c}
\includegraphics[height=8.0cm]{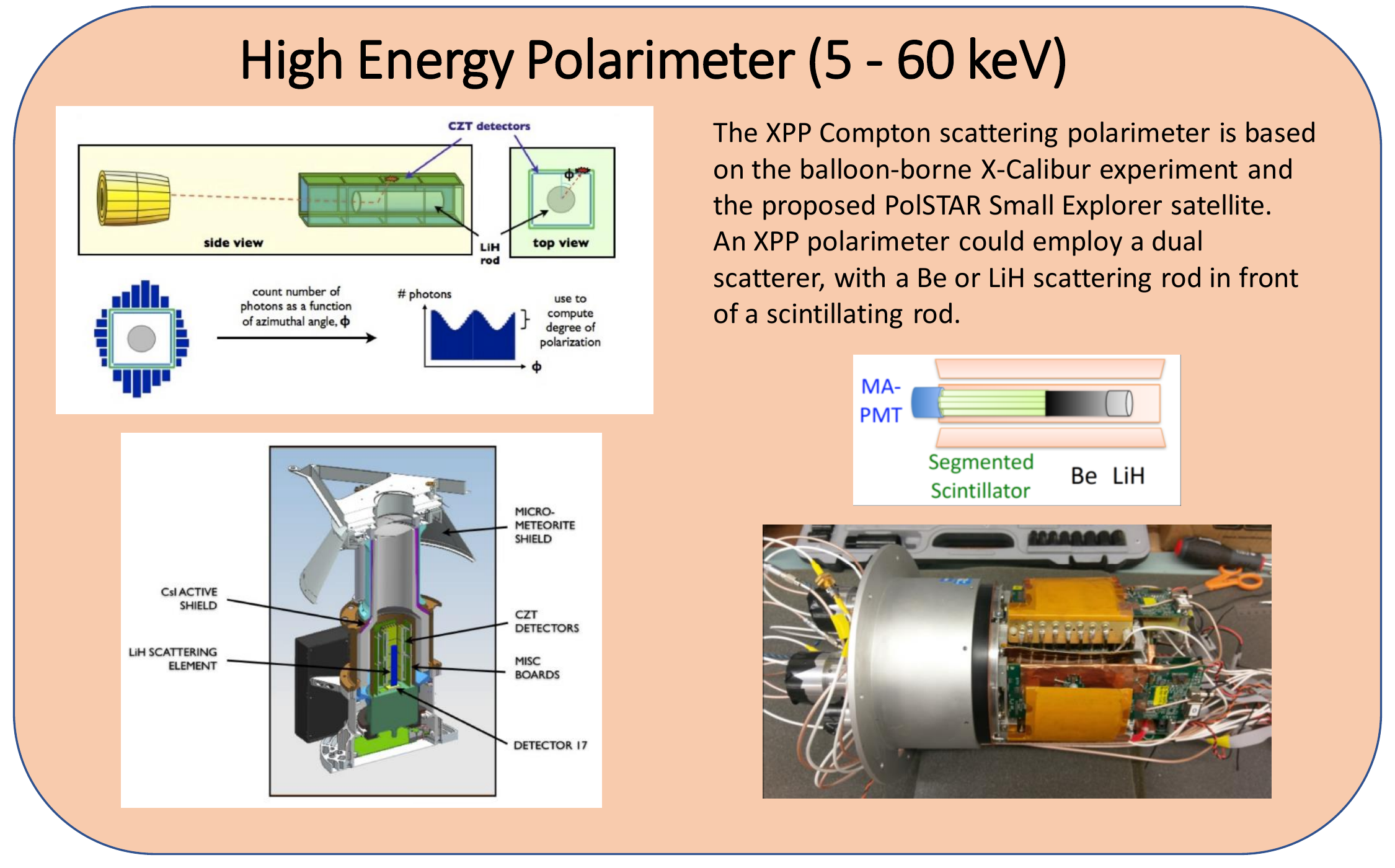}
\end{tabular}
\end{center}
\caption 
{ \label{fig:HEP}
The High Energy Polarimeter is a Compton Scattering polarimeter.   For {\it XPP}, the HEP would be just behind the focal plane and be illuminated by X-rays which pass through the MEP.} 
\end{figure}

\subsubsection{Medium Energy Polarimeter}
The Medium Energy Polarimeter (MEP) exploits the polarization dependence of the photoelectric effect by measuring the initial direction of the photoelectron which is correlated with the photon electric field.    Detectors are based on the successful PRAXyS design \cite{Hill16}, although we assume that the same sensitivity can be achieved with a detector that has twice the pressure and half the depth (e.g. total quantum efficiency is conserved).    The individual tracks will be half as long.    Maintaining the same effective resolution along the track requires readout electrodes with a pitch of half that used for GEMS/PRAXyS.   Polarization sensitivity has been measured from 2 - 8 keV\cite{Iwakiri16} with modulation factors ranging from 20\% to 55\%.   Improvements in track reconstruction algorithms continue to improve the sensitivity\cite{Kitaguchi18}.
The PRAXyS concept also needs to be modified with a rear window so that high energy photons can exit the detector and interact in the Compton scattering experiment.

\subsubsection{High Energy Polarimeter}
The High Energy Polarimeter (HEP) exploits the angular dependence of the Compton scattering cross section which peaks 90$^\circ$ away from the electric field vector.    The active element consists of a cylinder of scattering material (with Lithium and Beryllium sections) surrounded by imaging detectors, similar to the {\it X-Calibur} and {\it PolSTAR} detectors \cite{Krawczynski16,Kislat18}.  A modulation factor of ${\sim}50\%$ over the full energy range of the HEP is achieved.  The recent 2018/2019 balloon flight of the {\it X-Calibur} mission resulted in the first constraints on the linear polarization of the 15-35 keV X-rays from the accreting strongly magnetized neutron star GX 301$-$2 and validated the detection principle and the technical implementation. The addition of a segmented scintillator rear scatterer would enable imaging polarimetry with $\sim$1 arcmin angular resolution.
\begin{figure}
\begin{center}
\begin{tabular}{c}
\includegraphics[height=8.0cm]{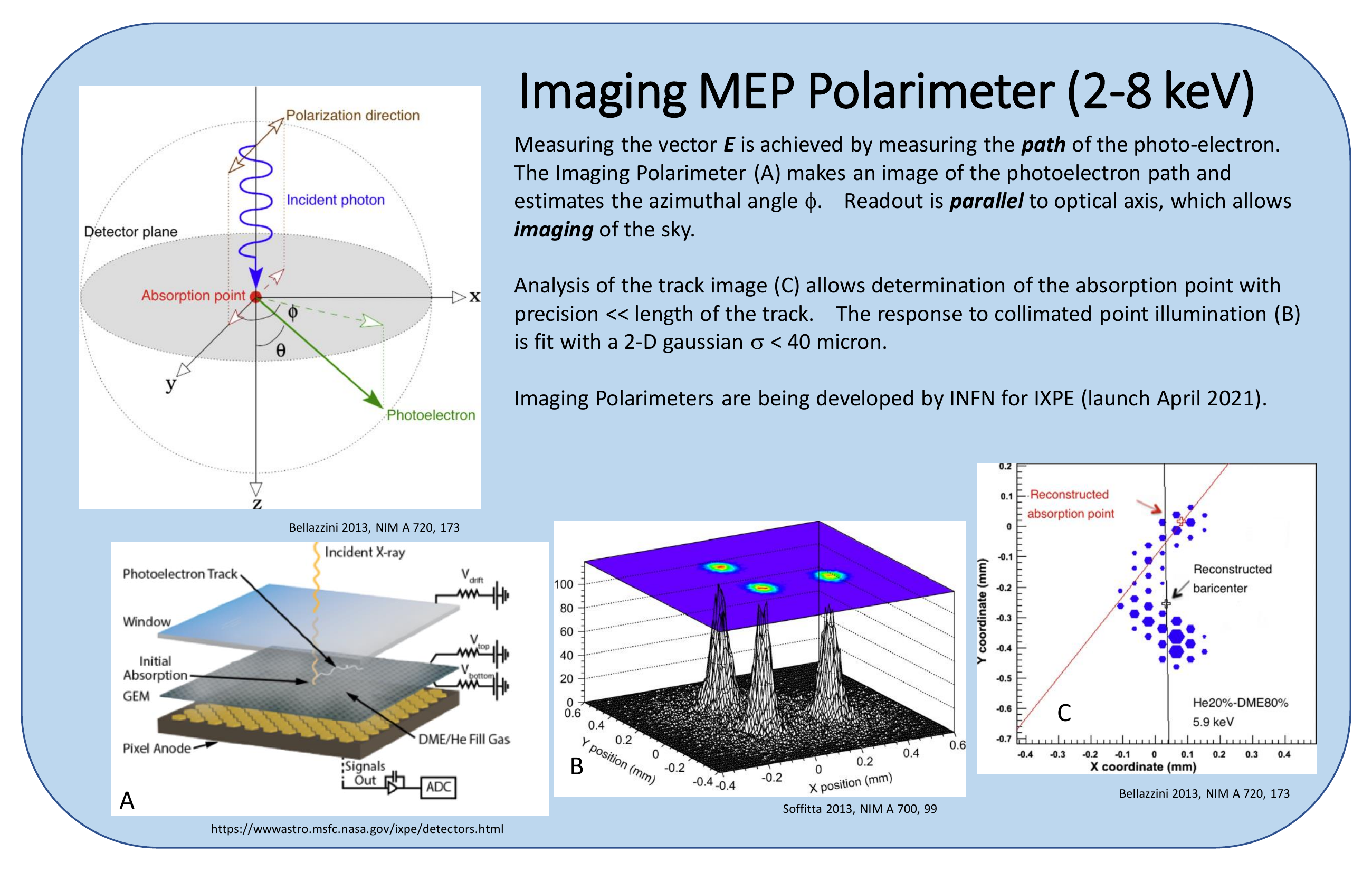}
\end{tabular}
\end{center}
\caption 
{ \label{fig:GPD}
The Imaging Medium Energy Polarimeter is based on the Gas Pixel Detectors being developed for IXPE.} 
\end{figure}

\subsubsection{Imaging Polarimeters}
The Imaging Polarimeter is a Gas Pixel Detector (GPD)\cite{2001Natur.411..662C}, developed in Italy and
refined significantly over the past 20 years.  As with the MEP, the goal is to measure the initial
direction of the photoelectron for 2-8 keV X-rays.  In the case of the GPD, however, the ionization
track drifts toward a Gas Electron Multiplier to generate more electrons for high signal/noise
and is then read out using an anode array with 50 $\mu$m hexagonal pixels.
{\it IXPE} will have three such detectors at the focal planes of three independent optics, so these
detectors will have a high technical readiness level after fabrication and calibration
this year and launch in 2021.
For {\it IXPE}, the anodes are directly connected to
CMOS-based Application Specific Integrated Circuits (ASICs)
with $\sim 10^5$ pixels with individual analog electronics\cite{2006NIMPA.566..552B}.
A schematic of the {\it IXPE} GPD is shown in Figure ~\ref{fig:GPD}.
The GPD spatial resolution is about 120 $\mu$m, so that the detector
can be paired with optics capable of 2-5" imaging, up to a factor of 10
improvement over {\it IXPE}.
The {\it IXPE} detectors are 1 cm deep, filled with a mixture of dimethyl ether and He at an 4:1 partial
pressure ratio, sealed at one atmosphere, and have shown no changes in performance
greater than 1\% per year.
Modulation factors of 20\% (2 keV) to 65\% (8 keV) have been measured in the lab\cite{2010NIMPA.620..285M}.
The bandpass of GPD (and TPC) polarimeters can be shifted and extended to higher energies (e.g.\ 6-35 keV) with suitable choices of the gas mixture and pressure (e.g.\ \cite{Tagliaferri11}). Using a single imaging polarimeter sensitive at higher energies, or replacing one of the scattering polarimeters with a high-energy GPDM might be interesting options that should be evaluated with suitable trade off studies.       

\subsection{Mission and Spacecraft components}

\subsubsection{Orbit and Launch Vehicle}
\begin{figure}
\begin{center}
\begin{tabular}{c}
\includegraphics[width=11cm]{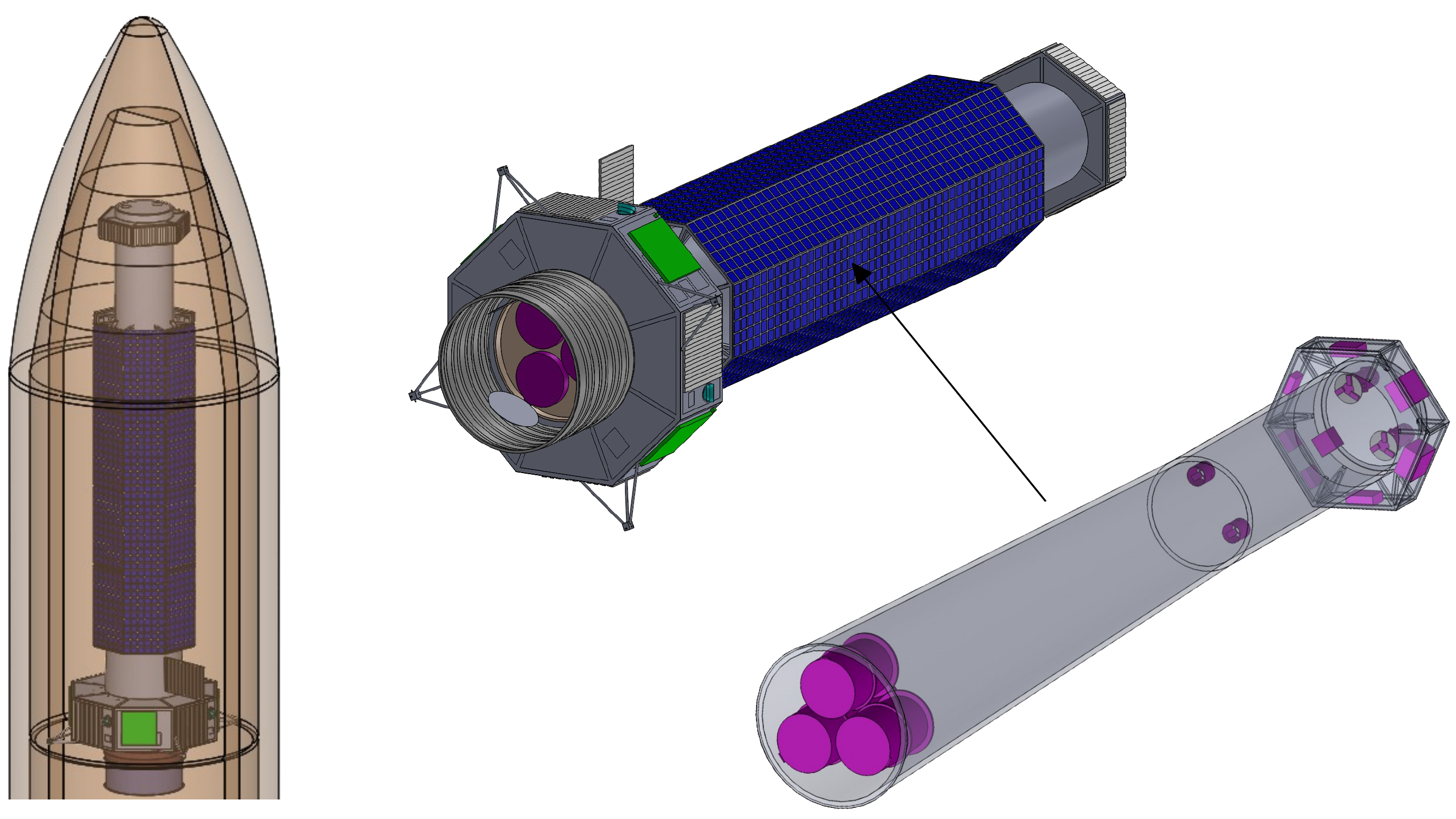}
\end{tabular}
\end{center}
\caption 
{ \label{fig:fairing}
Left: {\it{XPP}} inside the Atlas 5 meter Medium Payload Fairing.  Center:  {\it{XPP}} Observatory.  Right:  Payload structure (Observatory components removed).} 
\end{figure}

A medium class launch vehicle (e.g. Atlas 511) can place {\it XPP} directly into a 7000 x 114,000 km orbit.    This orbit has a 48 hour period, long life, and provides conditions favorable for observations for over 40 hours per orbit.   Lunar perturbations are important on long time scales;  apogee, perigee, and inclination vary on $\approx 10$ year time scales for the orbit studied here, with evolution similar to the {\it Chandra} orbit.   Figure \ref{fig:fairing} shows the Observatory inside the Atlas 5m fairing.    Only the sunshade and the thermal pre-collimators in front of the mirrors require deployment after launch.   The entire observatory rotates about the telescopes' line of sight with a period of $\approx$ 1 hour.

\subsubsection{Attitude Control}
The attitude control control system consists of currently and commercially available components including a Northrup Grumman internally redundant SIRU gyroscope, six coarse sun sensors, three SODERN SED36 star trackers (2 operational and a cold spare), six Honeywell HR-14 reaction wheels, reaction wheel isolators, an aspect camera coaligned with the telescope, and cold gas thrusters.    The star trackers provide 1 arcsec, 3$\sigma$ end of life (3 year) performance;  the reaction wheels can store up to 50 Nms of angular momentum and slew the spacecraft through 90 deg in 30 minutes.   Momentum unloading is performed with the cold gas thrusters approximately every 2 days, preferentially during the 8 hours surrounding perigee which are unsuited to observation.

\subsubsection{Propulsion}
Propulsion is required only to null out tip-off rates and for momentum management;  the launch vehicle is capable of delivering {\it{XPP}} to the final orbit.   60 kg of cold gas is estimated to provide enough capability to manage momentum for over 20 years.   Redundancy is provided by storing the gas in four tanks, each of which serves an independent Reaction Control System pod with 4 thrusters.    The pods are distributed uniformly in azimuth around the spacecraft.

\subsubsection{Mechanical}

The {\it{XPP}} primary structure combines instrument and spacecraft functions.   The telescope metering structure consists of a conical tube constructed from aluminum honeycomb and composite facesheets.   Mirrors are mounted within the larger end of the telescope tube;  detectors are mounted within the smaller end.    Spacecraft components are largely mounted outside the mirror end of the tube while instrument electronics and thermal control systems are mounted outside the detector end of the tube.  The outside of the tube is surrounded by fixed solar panels;  four phased array antenna panels surround the satellite bus.   As the antennas and solar panels are fixed, neither system is a source of disturbance torques to the pointing and control systems.   The rigid structure is continuously heated to remove the possibility of thermal distortions related to the slow rotation.

\subsubsection{Power and thermal control}
The fixed solar cell arrays provide sufficient power to operate in all orientations excluding 30 degree (half angle) cones about the sun and anti-sun directions.   Solar cell efficiency of 29\% and a three year degradation of 13\% are assumed.   Additional degradation (e.g. as might be expected for operations beyond 3 years) can be accommodated by increasing the size of solar keep out zones, or mission planning that ensures the battery charge will be high during potential solar occultations at perigee.
%

\subsubsection{Communications and Tracking}
{\it XPP} has four fixed phased array X-band antennas, decoupling the downlink schedule from observations.  A single DSN downlink with an 11 m  dish per day can downlink 100 Gb per day.   S band communications are used for commanding and some housekeeping.


\section {Technical Resources and Margins}
The {\it{XPP}} dry mass is estimated to be 2152 kg.   Allowing for 60 kg of cold gas propellant, and 40\% growth, a launch mass of 2904 kg was assumed for launch services calculations.   A 9\% margin to the Atlas 511 capability remains, even after derating the Atlas capability by 10\% and accounting for the mass of the adapter.

The solar arrays are designed to accommodate the highest power mode (Science plus communication) at the worst case solar angle (30 degrees between line of sight and the sun) after allowing for a 30\% growth in power demand.  Additional 'margin' can be created by reducing the field of regard.    {\it{XPP}} is not limited by considerations of mass or power.

Cold gas propellant is the only consumable on {\it{XPP}}.  For the current design, the propellant has an estimated lifetime of 20 years.

\section{Technology Drivers}
The {\it XPP} payload is based on instruments and components which substantial  design and test heritage. The spacecraft design presented here uses components and approaches available today.

\section{Organizations, Partnerships, and Current Status}
The {\it XPP} study was performed by representatives from MIT, GSFC and Washington University with input from {\it IXPE} members at the MSFC. These individuals and institutions have been building and qualifying versions of the LEP, MEP, HEP, and {\it IXPE} for a number of years and expect to work collaboratively to propose the {\it XPP} when the opportunity arises. We expect substantial international contributions from Italian and Japanese co-investigators with regards to the mirror, the polarimeters, the analysis, and scientific participation and leadership.  

\clearpage
\section{Cost and Schedule}
\label{sect:cost}
\begin{wrapfigure}{R}{0.5\textwidth}
\vspace*{-4ex}
\begin{center}
\begin{tabular}{c}
\includegraphics[width=0.48\textwidth]{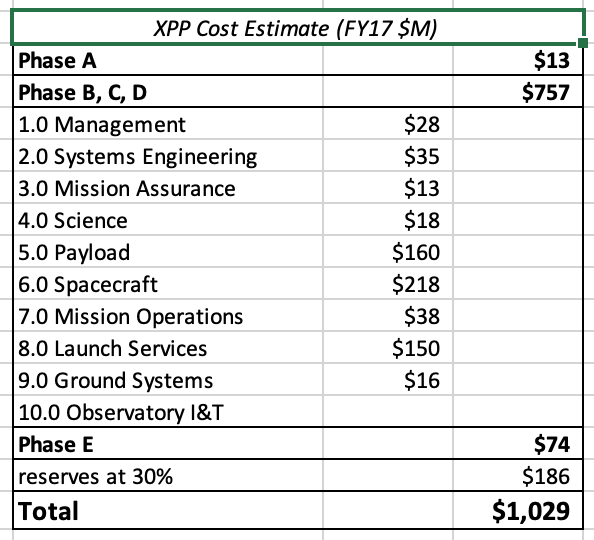}
\end{tabular}
\end{center}
\caption 
{ \label{fig:costs}
{\it XPP} cost estimate summarized by WBS.} 
\end{wrapfigure}
A mission cost estimate was developed as part of a Concept study performed at the Glenn Research Center COMPASS design center in September 2017.  The Life Cycle Cost includes 3 years of operations.   The {\it XPP} investigators self-assessment that the payload technologies are at TRL 6 is based on the on-going development for flight of the {\it IXPE} detectors, the flight design and independent TRL 6 qualification of the GEMS/PRAXyS detectors, the successful development of the {\it X-Calibur} balloon detectors, and the advanced development of all key components of the LEP.   

Payload costs (WBS 5) were estimated using PRICE True Planning and payload component masses estimated by the investigators.

Spacecraft costs (WBS 6) are based on the assumption that the spacecraft is provided by an industry partner responsible for all components (no GFE) and a 10\% fee is included.

Because payload components are distributed throughout the Observatory, Integration and Test costs were estimated for the combined Payload and Spacecraft system.  30\% of these costs are included in the spacecraft costs (WBS 6) with the balance accounted for in the traditional NASA WBS 10 (Observatory I\&T) element.

The total Phase E costs were estimated from the Mission Operations Cost Estimating Tool (MOCET) and include (parametrically) the downlink costs (~\$5M for a 3 year mission) and some Science Operations in addition to the costs traditionally carried in WBS 7.

The Launch costs (WBS 8) were taken to be \$150M, which was the guidance being given to Probe Class studies being performed in the GSFC Mission Design Lab at the time the {\it XPP} study was performed.    Reserves are not carried against this cost.

A risk analysis based on a Monte Carlo analysis which incorporated the relative uncertainties of each element of the costs produced a point estimate which was equal to the mean of all the runs.   All costs are presented in constant FY17 dollars.

The cost estimate is summarized by NASA standard top level WBS elements in Figure \ref{fig:costs}.

\section{Summary}
X-ray polarimetry is now becoming an observational area of astrophysics. A second-generation instrument like {\it XPP}  will vastly increase the number of observable sources and will be able to probe even small polarization fractions. Imaging polarimetry and broad-band spectro-polarimetry will provide qualitatively new insights into the structure of a broad range of astrophysical objects and have the power to revolutionize our understanding of many of the most energetic phenomena in the Universe.

\clearpage
\acknowledgments 
We gratefully acknowledge the COMPASS Mission Design Team at the NASA Glenn Research Center:  Steve Oleson, Tony Colozza, James Fincannon, James Fittje, John Gyekenyesi, Robert Jones, Nicholas Lantz,  Mike Martini, John Mudry, J. Michael Newman, Tom Packard, Dave Smith,  Sarah Tipler,  and Elizabeth Turnbull.  


\bibliography{report}   
\bibliographystyle{spiejour}   


\end{spacing}
\end{document}